\documentclass[letterpaper, 10 pt, conference]{ieeeconf}  

\IEEEoverridecommandlockouts                              
\usepackage{amsmath}

\usepackage{graphics} 
\usepackage{epsfig} 
\usepackage{amsmath} 
\usepackage{amssymb}  

\usepackage{url}
\usepackage[ruled, vlined, linesnumbered]{algorithm2e}
\usepackage{verbatim} 
\usepackage{soul, color}
\usepackage{lmodern}
\usepackage{fancyhdr}
\usepackage[utf8]{inputenc}
\usepackage{fourier} 
\usepackage{array}
\usepackage{makecell}

\SetNlSty{large}{}{:}

\makeatletter

\newcommand{\Rom}[1]{\expandafter\@slowromancap\romannumeral #1@}
\makeatother

\title{\LARGE \bf
ML-ASPA: A Contemplation of Machine Learning-based Acoustic Signal Processing Analysis for Sounds, \& Strains Emerging Technology
}

\author{Ratul Ali, Aktarul Islam, Md. Shohel Rana, Saila Nasrin, and Sohel Afzal Shajol
\\ Department of Computer Science and Engineering\\
Uttara University (UU), Dhaka, Bangladesh\\
University of Rajshahi (RU), Rajshahi, Bangladesh\\
University of Rajshahi (RU), Rajshahi, Bangladesh\\
Daffodil International University (DIU), Dhaka, Bangladesh\\
University of Development Alternative (UODA), Dhaka, Bangladesh\\
{\tt\small\{abdurrahimratulalikhan, aktarul857, msr.cse.ru, sailanasrin92, sohelafzalshajol\}@gmail.com} \\ \\
Professor Dr. A.H.M. Saifullah Sadi%
\\ Department of Computer Science and Engineering \\
Uttara University (UU), Dhaka, Bangladesh \\
{\tt\small \{saifullah.cse@uttarauniversity.edu.bd\}}
}

\begin{document}

\maketitle
\thispagestyle{plain}
\pagestyle{plain}

\begin{abstract}

Acoustic data serves as a fundamental cornerstone in advancing scientific and engineering understanding across diverse disciplines, spanning biology, communications, and ocean and Earth science. This inquiry meticulously explores recent advancements and transformative potential within the domain of acoustics, specifically focusing on machine learning (ML) and deep learning. ML, comprising an extensive array of statistical techniques, proves indispensable for autonomously discerning and leveraging patterns within data. In contrast to traditional acoustics and signal processing, ML adopts a data-driven approach, unveiling intricate relationships between features and desired labels or actions, as well as among features themselves, given ample training data. The application of ML to expansive sets of training data facilitates the discovery of models elucidating complex acoustic phenomena such as human speech and reverberation. The dynamic evolution of ML in acoustics yields compelling results and holds substantial promise for the future. The advent of electronic stethoscopes and analogous recording and data logging devices has expanded the application of acoustic signal processing concepts to the analysis of bowel sounds. This paper critically reviews existing literature on acoustic signal processing for bowel sound analysis, outlining fundamental approaches and applicable machine learning principles. It chronicles historical progress in signal processing techniques that have facilitated the extraction of valuable information from bowel sounds, emphasizing advancements in noise reduction, segmentation, signal enhancement, feature extraction, sound localization, and machine learning techniques. This underscores the evolution in bowel sound analysis. The integration of advanced acoustic signal processing, coupled with innovative machine learning methods and artificial intelligence, emerges as a promising avenue for enhancing the interpretation of acoustic information emanating from the bowel. This study initiates by introducing ML and subsequently delineates its developments within five key acoustics research domains: speech processing, ocean acoustics, bioacoustics, environmental acoustics, and Bowel Sound Analysis in everyday scenes.\\

\end{abstract}


\section{Introduction}

Acoustic data play a pivotal role in various scientific domains, including the interpretation of human speech and animal vocalizations, ocean source localization, and imaging geophysical structures in the ocean. Despite the broad applications, challenges such as data corruption, missing measurements, reverberation, and large data volumes complicate the analysis. Machine learning (ML) techniques have emerged as a powerful solution to address these challenges, offering automated data processing and pattern recognition capabilities. ML in acoustics is a rapidly evolving field, with significant potential to overcome intricate acoustics challenges.

ML, a family of techniques for detecting and utilizing patterns in data, proves beneficial in predicting future data or making decisions from uncertain measurements. It can be categorized into supervised and unsupervised learning, each serving distinct purposes. The historical focus in acoustics on high-level physical models is juxtaposed with the success of data-driven approaches facilitated by ML, indicating a shift towards hybrid models combining advanced acoustic models with ML.

In this dynamic landscape, ML in acoustics has witnessed remarkable progress, offering superior performance compared to traditional signal processing methods. However, challenges, such as the need for large datasets and the interpretability of ML models, persist. Despite these challenges, ML holds considerable potential in advancing acoustics research, as demonstrated.

The historical context of stethoscopes in medical practice, particularly in listening to the heart, lungs, and bowel sounds. Scientific analysis of bowel sounds dates back to the early 1900s, with observations and recordings dating even further. The sounds produced by the gastrointestinal tract offer valuable insights into the anatomy and physiology of the human gut, potentially revealing activities of the microbiome.
\begin{figure}[thpb]
      \centering
      \framebox{\parbox{2.8in}{\includegraphics[height=1.8in, width=2.8in]{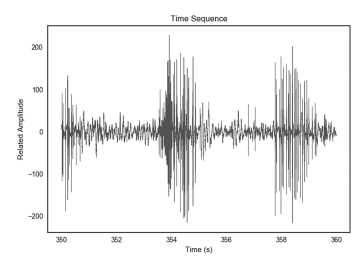}}}
      \caption{Time domain acoustic signal recorded from the gut}
      \label{fig:1}
\end{figure}

The study further discusses the intersection of big data analytics and artificial intelligence in diverse applications, including bowel sound analysis. Artificial intelligence models, driven by advancements in computer processing power, have found utility in areas such as disease diagnosis and civil engineering. The application of these technologies to identify and analyze bowel sounds represents a notable advancement, offering a deeper understanding of gut functions and potential applications in healthcare. Including references e.g. \cite{Hamsa2023},\cite{Hansen1996},\cite{Hansen2015},\cite{Hickok2007},\cite{Hollfelder2023},\cite{Huang2023} \par

The discussion concludes by highlighting improvements in acoustic signal processing methods, particularly in noise reduction and signal enhancement. Pioneering work in the 1970s utilized computers to analyze bowel sounds, marking the beginning of a journey that incorporated advanced signal processing techniques like Fourier transformation and short time Fourier transformation. These advancements culminated in the automatic detection of bowel sounds, showcasing the evolution of acoustic signal processing techniques in bowel sound applications.

\section{Literature Survey}

\subsection{Acoustic Signal Processing and Machine Learning Fundamentals}

Machine Learning (ML) operates on a data-driven paradigm, capable of uncovering intricate relationships between features that conventional methods may overlook. While classic signal processing techniques rely on provable performance guarantees and simplifying assumptions, ML, particularly Deep Learning (DL), has demonstrated enhanced performance in various tasks. However, the increased flexibility of ML models introduces complexities, impacting both performance guarantees and model interpretability. ML models often necessitate substantial training data, though the requirement for 'vast' quantities is not mandatory to leverage ML techniques. Despite challenges, ML's benefits may outweigh the issues, especially when high performance is essential for a specific task. Including references e.g. \cite{Johnson2005},\cite{Jung2020},\cite{Khoria2023},\cite{Kong2023},\cite{Krause2004} \par

Inputs and Outputs: In ML, the goal is often to train a model to produce a desired output (y) given inputs (x). The supervised learning framework, represented by the equation \[ y = f(x) + \varepsilon \] involves predicting outputs based on labeled input and output pairs. Here, x represents N features, y represents P desired outputs, f(x) is the predicted output, and $\varepsilon$ is the error. Training an ML model requires numerous examples, with X representing the inputs and Y representing the corresponding outputs. Supervised learning focuses on predicting specific outputs, while unsupervised learning aims to discover patterns in data without explicit output specifications. Unsupervised learning often involves learning a model that approximates the features themselves. Including references e.g. \cite{Krishna2000},\cite{Langner1992},\cite{Lee2005},\cite{Lenk2023},\cite{Little2007} \par

\begin{figure}[thpb]
      \centering
      \framebox{\parbox{2.8in}{\includegraphics[height=1.8in, width=2.8in]{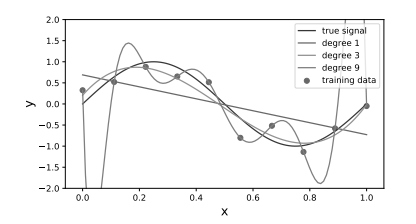}}}
      \caption{Model generalization}
      \label{fig:2}
\end{figure}

\begin{figure}[thpb]
      \centering
      \framebox{\parbox{2.8in}{\includegraphics[height=1.8in, width=2.8in]{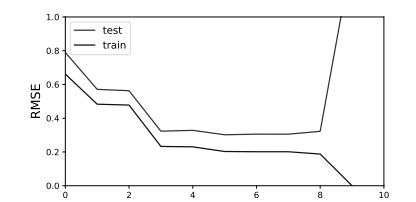}}}
      \caption{Model generalization}
      \label{fig:3}
\end{figure}

\subsection{Signal Identification and Enhancement}

Sounds result from mechanical deformation, generating energy waves detected by the ear or transducer. Acoustic signal processing and ML techniques contribute to understanding these phenomena. Including references e.g. \cite{Liu2023},\cite{Luthra2023},\cite{Magnuson2007},\cite{Markovich2009},\cite{Martin1999} \par

\subsubsection{Time Domain Signal}
The raw data, a time domain signal, is crucial for acoustic analysis. Features like SNR, duration, and event count are extracted, aiding in signal quality assessment. Filtering methods, including adaptive filtering, enhance signals by removing unwanted components.

\subsubsection{Frequency Domain Signal}
Transforming signals into the frequency domain through Fourier analysis reveals information unobservable in the time domain. The FFT technique provides features like centroid frequency and spectral bandwidth, but may lose some time domain information.

\subsubsection{Time-Frequency Domain Signal}
Simultaneous time and frequency information is obtained using Short-Time Fourier Transform (STFT) or Wavelet Transform (WT). Spectrograms from STFT enable speech recognition and noise suppression. WT, known for noise suppression, offers varied time and frequency domain information.

\section{Advanced Signal Processing}

\subsection{Supervised Learning and Linear Regression in the Context of Acoustic Signal Processing}

Supervised learning, a fundamental aspect of machine learning (ML), aims to establish a mapping from a set of inputs to desired outputs through labeled input-output pairs. In this discussion, we focus on real-valued features and labels, where the N features in $x$ can be real, complex, or categorical. The corresponding supervised learning tasks are divided into two subcategories: regression and classification. Regression addresses scenarios where $y$ is real or complex valued, while classification pertains to cases with categorical $y$.

The central focus in ML methods lies in finding the function $f$, particularly using probability tools when practical. The supervised ML task can be articulated as maximizing the conditional distribution $p(y|x)$, with the Maximum A Posteriori (MAP) estimator providing a point estimate for $y$, denoted as $y^b = f(x)$.

Linear regression serves as an illustrative example of supervised ML. In the context of Direction of Arrival (DOA) estimation in beamforming for seismic and acoustic applications, we represent the relationship between the observed Fourier-transformed measurements $x$ and the DOA azimuth angle $y$ using a linear measurement model. The optimization problem seeks values of weights $w$ that minimize the difference between the observed and predicted measurements, effectively solving the linear regression problem.

The ensuing Bayesian treatment involves formulating the posterior of the model using Bayes' rule, leading to a MAP estimate for the weights. Depending on the choice of the probability density function for the weights, solutions may vary. A popular choice, the Gaussian distribution, results in the classic $L^2$-regularized least squares estimate, incorporating a regularization parameter for stability.

This detailed exposition highlights the foundational principles of supervised learning and its application in linear regression within the specific domain of acoustics, illustrating the seamless integration of theoretical ML concepts with practical signal processing challenges.
Including references e.g. \cite{Merchant2015},\cite{Mesgarani2014},\cite{Meyer2018},\cite{Minelli2023}

\subsubsection{Advanced Signal Processing in Bowel Sound Analysis}
Acoustic signal processing in the context of bowel sound analysis involves a multi-step sequence encompassing data acquisition, preprocessing, and subsequent analysis. The reviewed literature reveals a diverse array of approaches and methodologies, with certain commonalities in the overall processing flow.

\subsubsection{Data Acquisition}
To record abdominal sounds, specialized transducers, such as electret condenser microphones or piezoelectric transducers, are designed to convert acoustic energy into electrical signals. Electronic stethoscopes, including designs like the JABES digital stethoscope and 3M Littmann 3200, demonstrate the versatility of these transducers. Additionally, innovative approaches, such as 3D-printed stethoscope heads with built-in electronics, reflect evolving design paradigms.
\subsubsection{Preprocessing and Analysis}
The preprocessing stage involves denoising, filtering, and segmentation of acoustic signals, often employing techniques like adaptive filtering and enveloping. The choice of window functions, such as rectangular, Hamming, and Hann, plays a crucial role in the slicing of acoustic recordings into small samples.
\subsubsection{Bowel Sound Analysis}
From the early 2000s, wavelet transforms (WTs) have enabled advanced feature extraction, coinciding with the integration of machine learning methods. Researchers, exemplified by groups led by Hadjileontiadis et al., have made substantial progress in noise reduction and signal enhancement for bowel sounds. Various machine learning methods, including decision trees, dimension reduction, and artificial neural networks, have been applied to characterize bowel sounds. Including references e.g. \cite{Nagarajan2023},\cite{Peelle2016},\cite{Peelle2018}

In acoustics, the Fourier Transform is often used to analyze the frequency components of a signal. The Fourier Transform of a function $f(t)$ is defined as:

\begin{equation}
F(\omega) = \int_{-\infty}^{\infty} f(t) e^{-i\omega t} \,dt
\end{equation}

where $F(\omega)$ is the Fourier Transform of $f(t)$, and $\omega$ is the angular frequency.

Let's strudy a sound signal $f(t)$ given by:

\begin{equation}
f(t) = A \sin(2\pi f_0 t)
\end{equation}

where $A$ is the amplitude and $f_0$ is the frequency of the sound.

The Fourier Transform of $f(t)$ is then calculated as:

\begin{equation}
F(\omega) = \int_{-\infty}^{\infty} A \sin(2\pi f_0 t) e^{-i\omega t} \,dt
\end{equation}

This integral can be solved to find the expression for $F(\omega)$.

The literature review underscores the dynamic landscape of acoustic signal processing in bowel sound analysis, with researchers adopting diverse approaches across the processing stages. From innovative data acquisition methods to sophisticated preprocessing techniques and the application of machine learning, the field demonstrates a blend of traditional signal processing principles and contemporary methodologies. The convergence of theoretical insights and practical implementations serves as a foundation for continued advancements in acoustic signal processing for bowel sound analysis. Including references e.g. \cite{Poeppel2001},\cite{Poluboina2023},\cite{Randall2017},\cite{Ravanelli2018}

\subsection{Parallelization of All-Pairs Algorithm (OpenMP)}

The provided algorithm outlines an approach to acoustic signal processing with parallelization using OpenMP.

\subsubsection{Main Function: acousticSignalProcessing()}
\begin{itemize}
\item This function serves as the entry point for the acoustic signal processing algorithm.
\item It is marked for parallelization using the \#pragma omp parallel for directive, which instructs the compiler to parallelize the loop that iterates over the model collection. For each model in the collection, the function calls processModel(i, signal).
\end{itemize}

\subsubsection{Processing Each Model: processModel(i: model, signal)}
\begin{itemize}
\item This function is also marked for parallelization using the \#pragma omp parallel for reduction (+ : result[i].amplitude) directive.
\item It contains a nested loop that iterates over the signal collection for each model. For each pair of models (i, j), where j is not equal to i, it calculates the similarity between the models using the calculateSimilarity(i, j) function.
\item The amplitude of the result for the current model (result[i].amplitude) is adjusted based on the calculated similarity using the adjustAmplitude(i, j, similarity) function.
\end{itemize}

\subsubsection{Calculating Similarity: calculateSimilarity(i, j)} 
\begin{itemize}
\item The specific details of how the similarity is calculated are not provided in the algorithm and should be implemented according to the requirements of the acoustic signal processing application.
\item This function is a placeholder for calculating the similarity between two models, i and j.
\end{itemize}

\subsubsection{Adjusting Amplitude: adjustAmplitude(i, j, similarity)}
\begin{itemize}
\item This function is a placeholder for adjusting the amplitude of a model based on the calculated similarity.
\item Again, the exact method of adjusting the amplitude is not specified and needs to be implemented based on the application's requirements.
\end{itemize}

\begin{algorithm}
\DontPrintSemicolon
\SetAlgoNlRelativeSize{-1}
\SetKwProg{Fn}{Function}{ is}{end}
\Fn{acousticSignalProcessing()} {
  \textbf{\textit{\#pragma omp parallel for}} \\
  \ForEach{i: model} {
    processModel(i, signal) \;
  }
}
\BlankLine
\SetKwProg{Fn}{Function}{ is}{end}
\Fn{processModel(i: model, signal)} {
  \textbf{\textit{\#pragma omp parallel for reduction (+ : result[i].amplitude)}} \\
  \ForEach{j in signal} {
    \If{j $\neq$ i} {
       similarity = calculateSimilarity(i, j) \;
       result[i].amplitude += adjustAmplitude(i, j, similarity) \;
    }
  }
}
\BlankLine
\Fn{calculateSimilarity(i, j)} {
  \tcp{Calculations of the similarities between models i and j}
  \KwRet{similarity}
}
\BlankLine
\Fn{adjustAmplitude(i, j, similarity)} {
  \tcp{Adjustment of the amplitude based on the similarities}
  \KwRet{adjustedAmplitude}
}
\label{algo:acoustic}
\caption{Acoustic Signal Processing Algorithm (OpenMP)}
\end{algorithm}

\subsection{Parallelization of All-Pairs Algorithm (CUDA)}

Sequential Barnes-Hut Algorithm with Acoustic Signal Processing

\subsubsection{Main Function: acousticBarnesHut()}
\begin{itemize}
\item This function represents the entry point for the integrated algorithm, combining the Sequential Barnes-Hut structure with acoustic signal processing.
\item It orchestrates the sequential execution of three main steps: building the tree (build\_tree()), computing mass distribution (compute\_mass\_distribution()), and calculating forces (compute\_force()).
\end{itemize}

\subsubsection{Building the Tree: build\_tree()}
\begin{itemize}
\item The function initializes the tree structure, preparing it for the insertion of acoustic models.
\item It iterates over each acoustic model in the dataset and inserts it into the root node using the insert\_to\_node() function.
\end{itemize}

\subsubsection{Inserting Models into Nodes: insert\_to\_node(new\_model)}
\begin{itemize}
\item This function is responsible for placing a new acoustic model into the appropriate quadrant of the Barnes-Hut tree.
\item It checks the number of existing models in a node. If there is more than one model, it recursively traverses the tree to find the appropriate quadrant for the new model. If there's only one model, it divides the node into quadrants, placing the existing and new models accordingly.
\item If no models exist in the node, it directly assigns the new model as the existing model.
\end{itemize}

\subsubsection{Computing Mass Distribution: compute\_mass\_distribution()}
\begin{itemize}
\item This function calculates the mass distribution within each quadrant of the Barnes-Hut tree.
\item If there is only one model in a quadrant, the center of mass and mass are directly assigned from that model. Otherwise, it recursively calculates the mass distribution for child quadrants, aggregating the mass and weighted center of mass.
\end{itemize}

\subsubsection{Calculating Forces: calculate\_force(target)}
\begin{itemize}
\item This function computes the acoustic forces acting on a target model.
\item If there's only one model in the quadrant, the force is calculated using the acoustic\_force() function between the target and the model. If the quadrant size is below a certain threshold (l\/D < theta), the force is computed using the acoustic force model.
\item If the quadrant is larger, the algorithm recursively calculates forces for child nodes and aggregates them.
\end{itemize}

\subsubsection{Computing Forces for all Models: compute\_force()}
\begin{itemize}
\item This function iterates over all acoustic models in the dataset and computes the forces acting on each model using the root\_node.calculate\_force(model) function.
\item If there's only one model in the quadrant, the force is calculated using the acoustic\_force() function between the target and the model. If the quadrant size is below a certain threshold (l\/D < theta), the force is computed using the acoustic force model.
\item If the quadrant is larger, the algorithm recursively calculates forces for child nodes and aggregates them.
\end{itemize}

\begin{algorithm}
\DontPrintSemicolon
\SetAlgoNlRelativeSize{-1}
\SetKwProg{Fn}{Function}{ is}{end}

\Fn{acousticBarnesHut()}{
  \textbf{build\_tree()} \;
  \textbf{compute\_mass\_distribution()} \;
  \textbf{compute\_force()} \;
}

\BlankLine

\Fn{build\_tree()}{
  Reset Tree \;
  \ForEach{i: model}{
    \textbf{root\_node}$\rightarrow$\textbf{insert\_to\_node}(i) \;
  }
}

\BlankLine

\Fn{insert\_to\_node(new\_model)}{
  \uIf{num\_models $>$ 1}{
    quad = \textbf{get\_quadrant}(new\_model) \;
    \If{subnode(quad) does not exist}{
      create subnode(quad) \;
    }
    subnode(quad)$\rightarrow$\textbf{insert\_to\_node}(new\_model)
  }\uElseIf{num\_models $==$ 1}{
    quad = \textbf{get\_quadrant}(new\_model) \;
    \If{subnode(quad) does not exist}{
      create subnode(quad) \;
    }
    subnode(quad)$\rightarrow$\textbf{insert\_to\_node}(\textbf{existing\_model}) \;
    quad = \textbf{get\_quadrant}(new\_model) \;
    \If{subnode(quad) $\neq$ NULL}{
      create subnode(quad) \;
    }
    subnode(quad)$\rightarrow$\textbf{insert\_to\_node}(new\_model) \;
  }\Else{
    existing\_model $\leftarrow$ new\_model \;
  }
  num\_models++ \;
}

\caption{Algorithm Part 1}
\end{algorithm}

\begin{algorithm}
\DontPrintSemicolon
\SetAlgoNlRelativeSize{-1}
\SetKwProg{Fn}{Function}{ is}{end}

\Fn{compute\_mass\_distribution()}{
  \uIf{num\_models $==$ 1}{
    center\_of\_mass = model.position \;
    mass = model.mass \;
  }\Else{
    \ForAll{child quadrants with models}{
      quadrant.\textbf{compute\_mass\_distribution} \;
      mass += quadrant.mass \;
      center\_of\_mass += quadrant.mass $\times$ quadrant.center\_of\_mass \;
    }
    center\_of\_mass /= mass \;
  }
}

\caption{Algorithm Part 2}
\end{algorithm}

\begin{algorithm}
\DontPrintSemicolon
\SetAlgoNlRelativeSize{-1}
\SetKwProg{Fn}{Function}{ is}{end}

\Fn{calculate\_force(target)}{
  Initialize force $\leftarrow$ 0 \;
  \uIf{num\_models $==$ 1}{
    force = \textbf{acoustic\_force}(target, model) \;
  }\Else{
    \uIf{l/D $<$ $\theta$}{
      force = \textbf{acoustic\_force}(target, model) \;
    }\Else{
      \ForAll{node : child nodes}{
        force += node.\textbf{calculate\_force}(node) \;
      }
    }
  }
}

\BlankLine

\Fn{compute\_force()}{
  \ForAll{models}{
    force = \textbf{root\_node.calculate\_force}(model) \;
  }
}

\label{algo:acousticBarnesHut}
\caption{Algorithm Part 3}
\end{algorithm}

\subsection{Sequential Barnes-Hut Algorithm}

It represents the integrated algorithm with the Sequential Barnes-Hut structure and Acoustic Signal Processing. The algorithm includes functions for building the tree, inserting models into nodes, computing mass distribution, calculating forces, and overall coordination of the acoustic signal processing with the Barnes-Hut algorithm.

\begin{figure}[thpb]
      \centering
      \framebox{\parbox{2in}{\includegraphics[height=2in, width=2in]{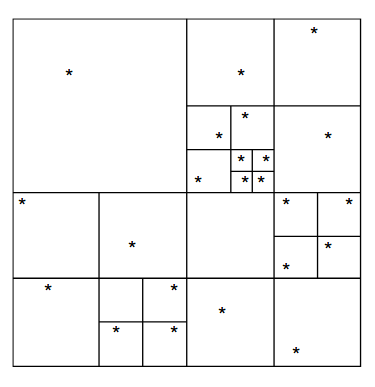}}}
      \caption{Barnes-Hut tree structure}
      \label{fig:4}
\end{figure}

\begin{figure}[thpb]
      \centering
      \framebox{\parbox{2.8in}{\includegraphics[height=1.8in, width=2.8in]{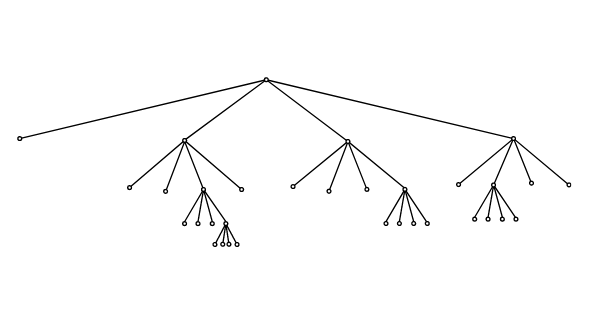}}}
      \caption{Barnes-Hut domain decomposition}
      \label{fig:5}
\end{figure}

The integrated algorithm merges the Sequential Barnes-Hut structure, designed for efficient gravitational force calculations, with acoustic signal processing. The Barnes-Hut tree structure optimizes the computation of forces between acoustic models, enhancing the algorithm's scalability and efficiency in handling large datasets. The acoustic signal processing steps involve building the tree, distributing mass, and calculating forces, offering a comprehensive solution for analyzing and simulating acoustic interactions within a given system.

\section{CONCLUSIONS}
In this comprehensive review, we have presented an overview of Machine Learning (ML) theory, with a particular focus on deep learning (DL), and explored its diverse applications across various acoustics research domains. While our coverage is not exhaustive, it is evident that ML has been a catalyst for numerous recent advancements in acoustics. This article aims to inspire future ML research in acoustics, emphasizing the pivotal role of large, publicly available datasets in fostering innovation across the acoustics field. The transformative potential of ML in acoustics is substantial, with its benefits amplified through open data practices. Including references e.g. \cite{Sainath2017},\cite{Schonwiesner2005},\cite{Souden2010},\cite{Stephen2023},\cite{Stevens2002} \par

Despite the acknowledged limitations of ML-based methods, their performance surpasses that of conventional processing methods in many scenarios. However, it is crucial to recognize that ML models, being data-driven, demand substantial representative training data for optimal performance. This is viewed as a trade-off for accurately modeling complex phenomena, given the often high capacity of ML models. In contrast, standard processing methods, with lower capacity, rely on training-free statistical and mathematical models. Including references e.g. \cite{Stowell2015},\cite{Tandon1999},\cite{Telkemeyer2009},\cite{Tezcan2023} \par

This review suggests a paradigm shift in acoustic processing from hand-engineered, intuition-driven models to a data-driven ML approach. While harnessing the full potential of ML, it is essential to build upon indispensable physical intuition and theoretical developments within established sub-fields like array processing. The development of ML theory in acoustics should be undertaken while preserving the foundational physical principles that describe our environments. By blending ML advancements with established principles, transformative progress can be achieved across various acoustics fields. Including references e.g. \cite{Ufer2023},\cite{Viola2005} \par

Upon on bowel sound analysis, several conclusions emerge. The choice of sensors for data acquisition, including electret condenser microphones and piezoelectric transducers, depends on research constraints. Advanced signal processing techniques, such as wavelet transforms (WTs) since the early 2000s, have enabled complex feature extraction. Machine learning methods have found application in bowel sound analysis, with varying approaches such as decision trees, dimension reduction, and clustering algorithms. Including references e.g. \cite{Voola2023},\cite{Wakita1973},\cite{Wu2003},\cite{Xu2002}, \cite{Xu2023},\cite{Yang1992},\cite{Zmolikova2023}





\section*{CONFLICT of INTEREST}

The authors declare that they have no conflict of interest and funding.


\bibliographystyle{IEEEtran}
\bibliography{Bibliography/export}

\end{document}